\documentclass[a4paper]{jpconf}
\usepackage{graphicx}
\begin{document}

\newcommand{\mean}[1]{\langle #1 \rangle} 

\title{Phase space constraints and statistical jet studies in heavy-ion collisions}

\author{Nicolas Borghini}

\address{Fakult\"at f\"ur Physik, Universit\"at Bielefeld, 
        Postfach 100131, D-33501 Bielefeld, Germany}

\ead{borghini@physik.uni-bielefeld.de}

\begin{abstract}
The effect of the correlation induced by global momentum conservation on the
two-particle distribution in nucleus--nucleus collisions is discussed, with a 
focus on the generic case of collisions with a non-vanishing impact parameter. 
\end{abstract}

\section{Introduction}
\label{s:intro}

As is well known, the distribution of particles in the final state of a 
collision between elementary particles results from the interplay of two 
ingredients~\cite{Yao:2006px}. 
First, there is a ``phase space'' part, which reflects the kinematic constraints
arising from energy and momentum conservation. 
Then, there comes the dynamical component, namely the detailed interaction(s) 
between the ``initial-state'' particles. 
This is arguably the most interesting part, inasmuch as one does not expect any
violation of energy-momentum conservation, where the modeling enters --- either 
first-principle based modeling for the interaction of two elementary particles, 
or more effective descriptions as required when more particles are involved ---
together with its set of assumptions and parameters. 
In particular, the aim of experimental investigations is to determine, or at 
least to constrain, those possible models, taking the conservation of energy 
and momentum as granted. 
For that purpose, it is necessary to have a good control of the ``physically 
trivial'' kinematic component. 

For single-particle distributions, the effect of energy-momentum conservation 
is typically to limit the value of the particle momentum and to result in a 
depletion in the yield close to the boundary: the effect amounts to a 
multiplicative factor. 
In the case of the joint distributions of two, three or more particles, the 
influence of the phase-space constraint is usually less straightforward. 
If there are only a few final-state particles, the joint distributions are 
quite constrained and the corresponding possible final states are conveniently 
represented on Dalitz plots~\cite{Yao:2006px}. 
However, in high-energy collisions, in particular of heavy nuclei, where many 
particles are emitted, such an approach becomes impractical. 
One has instead to estimate the effect of total momentum conservation on a 
statistical basis, be it for the induced correlation between two~\cite{%
  Foster:1973mt} or more than two~\cite{Borghini:2003ur} particles. 

Here, I shall briefly recall how the multiparticle correlation induced by the
conservation of total momentum can be derived in the limit of a large number of 
emitted particles~\cite{Borghini:2003ur}.
Then I shall focus on a case of interest for heavy-ion collision studies, 
namely when the emission of particles in the plane transverse to the beam is 
anisotropic (i.e., in the presence of so-called ``anisotropic flow'').

\section{Cumulants from momentum conservation}
\label{s:cor}

Let us consider a collision in which $N$ particles are emitted, viewed in the 
center-of-mass frame of the particles, so that the sum of their momenta vanishes
${\bf p}_1+\cdots+{\bf p}_N={\bf 0}$. 
This relationship induces a correlation between the momenta of $M$ particles 
chosen arbitrarily among the $N$ ones: technically, the corresponding joint 
$M$-particle probability distribution does not factorize into the product of the
$M$ single-particle probability distributions, but involves non-vanishing 
{\em cumulants\/}~\cite{vanKampen} of all orders.\footnote{Using probability 
  distributions, instead of the distributions themselves, is much simpler, for 
  it circumvents the issue of choosing a ``proper'' normalization --- e.g.\ of 
  particle-pair yields --- when the multiplicity $N$ fluctuates from event to 
  event (see the discussion in reference~\cite{Borghini:2004ra}, section~III).
  Yet, one should not forget that the problem is always present in a real 
  experimental correlation measurement.}

The strength of the $M$-particle correlation depends on both $M$ and the total 
number of particles $N$. 
In the simplest case of $N=2$ final-state particles, the correlation is maximal,
since ${\bf p}_2=-{\bf p}_1$; the corresponding two-particle probability 
distribution $f({\bf p}_1,{\bf p}_2)=f({\bf p}_1)\delta({\bf p}_1+{\bf p}_2)$ is 
obviously not factorizable. 
In the large-$N$ limit, one can prove with the help of a saddle-point 
approximation that the $M$-particle cumulant scales like $1/N^{M-1}$~\cite{%
  Borghini:2003ur}. 
A convenient and systematic approach to show that, as well as to obtain the 
explicit expressions of the cumulants, consists in using a generating function 
of the joint multiparticle distributions, the logarithm of which generates the 
cumulants. 
A few steps allow one to write this generating function as the integral of the 
exponential of $N$ times some function ${\cal F}({\bf k})$ of the integration 
variable ${\bf k}$. 
The procedure to derive the successive cumulants then consists in computing to 
a given order in powers of $1/N$ the position of the maximum of ${\cal F}$, 
i.e.\ the saddle point, then to calculate the value of ${\cal F}$ at this 
maximum so as to perform the saddle-point integration~\cite{Borghini:2003ur}. 

Admittedly, the cumulants induced by total momentum conservation are in general 
small when $N$ is large. 
However it is worth keeping their existence in mind, since they can become 
significant in some regions of phase space. 
At the same time, the correlations induced by other more dynamical phenomena 
which one attempts to investigate might conversely be small, and thus not 
necessarily significantly larger than the ``trivial'' kinematic ones. 
Thus, it has been argued that some influence of momentum conservation on the 
measured two-particle short-range correlations of identical pions in $pp$ 
collisions at RHIC energies can be evidenced~\cite{Chajecki:2006hn}. 
In heavy ion collisions, where $N$ is larger, momentum-conservation induced
correlations are even smaller, yet they could play a non-negligible role in 
some studies of small signals. 
Since it is important to have a good idea of what their effect might look like, 
so as to try to identify and subtract similar patterns in the measured 
correlations, I shall now discuss further these cumulants. 

\section{Azimuthally-dependent cumulants and distributions}

The most general expressions of the two- and three-particle cumulants due to 
the conservation of global momentum to leading order in powers of $1/N$, derived
according to the method sketched above~\cite{Borghini:2003ur}, can be found in 
reference~\cite{Borghini:2007ku} (equations (3.4) and (3.5)).
Neglecting the components of the momenta along the beam direction\footnote{%
  In the absence of an experimental estimate of the mean square momentum along 
  the beam as compared to along the transverse directions, this approximation 
  can tentatively be justified by the reported quasi-independence of the average
  transverse momentum $\mean{p_T}$ on rapidity~\cite{Bearden:2004yx}.},  
I shall focus on the constraint from {\em transverse\/} momentum conservation.

In the heavy-ion context, the mean square momenta along the nucleus--nucleus 
impact parameter (i.e., in the reaction plane) $\mean{p_x^2}$ and perpendicular 
to it in the transverse plane, $\mean{p_y^2}$, are generally unequal: this is 
the celebrated anisotropic expansion. 
To account for the phenomenon, let me introduce the coefficient
$\bar v_2\equiv\mean{p_x^2-p_y^2}/\mean{p_x^2+p_y^2}$.\footnote{This definition 
  differs from that of the usual elliptic-flow coefficient
  $v_2\equiv\mean{(p_x^2-p_y^2)/(p_x^2+p_y^2)}$, yielding values larger by 
  about a factor 2.}
Note that while a fourth-harmonic modulation $v_4$ of the single-particle 
distribution has also been evidenced at RHIC, I need not consider it here, as 
it would affect the cumulants induced by momentum conservation only from the 
four-particle cumulant onwards. 
Introducing $\bar v_2$ into the two-particle cumulant expression, one can recast
it as
\begin{equation}
\label{f_c(2)}
\bar f_c({{\bf p}_T}_1,{{\bf p}_T}_2) = 
-\frac{2{p_T}_1{p_T}_2}{N\mean{p_T^2}(1-\bar v_2^2)} \left[ 
  \cos(\varphi_2-\varphi_1) - \bar v_2\cos(\varphi_1+\varphi_2-2\Phi_R)\right],
\end{equation}
where $\Phi_R$ is the reaction-plane azimuth.
This two-particle cumulant depends not only on the relative angle 
$\Delta\varphi_{12}\equiv\varphi_2-\varphi_1$ between the particles, but also 
on the absolute orientation of the particle pair with respect to the reaction 
plane. 
In turn, the joint probability distribution $f({{\bf p}_T}_1,{{\bf p}_T}_2) = 
f(\varphi^{\rm pair},{p_T}_1,{p_T}_2,\Delta\varphi_{12})$ depends on the ``pair 
angle'' $\varphi^{\rm pair}\equiv(\varphi_1+\varphi_2)/2$ as well. 
Given equation~(\ref{f_c(2)}), computing the two-particle distribution is 
straightforward. 
Keeping only the second-harmonic modulation (``elliptic flow'') of the 
single-particle distributions, one finds
\begin{equation}
\label{f(2)}
f(\varphi^{\rm pair},{p_T}_1,{p_T}_2,\Delta\varphi_{12}) =
\frac{1}{2\pi}\left[ 1 + 
2v_{2,c}^{\rm pair} \cos 2(\varphi^{\rm pair}\!-\Phi_R) +
2v_{2,s}^{\rm pair} \sin 2(\varphi^{\rm pair}\!-\Phi_R) + \cdots\right],
\end{equation}
where the Fourier ``pair-flow'' coefficients~\cite{Borghini:2004ra} that 
characterise the azimuthal dependence of $f({{\bf p}_T}_1,{{\bf p}_T}_2)$ depend on 
${p_T}_1$, ${p_T}_2$ and $\Delta\varphi_{12}$. 
Considering only terms up to ${\cal O}(v_2/N)$ and ${\cal O}(v_2^3)$:\footnote{%
  This truncation of the expansion is driven by the respective values of $v_2$ 
  and $N$ ar SPS and RHIC, and is by no means mandatory.}
\begin{eqnarray}
\fl\displaystyle v_{2,c}^{\rm pair}({p_T}_1,{p_T}_2,\Delta\varphi_{12}) 
&\!\!\!\!\simeq\!\!\!& 
\left[1-2v_2(1)v_2(2)\cos(4\Delta\varphi_{12})\right]
\left[v_2(1)+v_2(2)\right]\cos(2\Delta\varphi_{12}) +
\frac{{p_T}_1{p_T}_2}{N\mean{p_T^2}}\bar v_2,\quad \label{v2c(2p)}\\
\displaystyle v_{2,s}^{\rm pair}({p_T}_1,{p_T}_2,\Delta\varphi_{12}) 
&\!\!\!\!\simeq\!\!\!& 
\left[1-2v_2(1)v_2(2)\cos(4\Delta\varphi_{12})\right]
\left[v_2(1)-v_2(2)\right]\sin(2\Delta\varphi_{12}), \label{v2s(2p)}
\end{eqnarray}
where $v_2(1)$, $v_2(2)$ are shorthand notations for $v_2({p_T}_1)$ and 
$v_2({p_T}_2)$, respectively. 
There are also higher harmonic terms ($v_{4,c}^{\rm pair}$, 
$v_{4,s}^{\rm pair}$\ldots), which are however smaller by at least a factor $v_2$. 
The effect of momentum conservation is actually subleading in the sine 
term~(\ref{v2s(2p)}), which reflects the non-invariance of the system under the 
$\varphi^{\rm pair}\!-\Phi_R\to -(\varphi^{\rm pair}\!-\Phi_R)$ symmetry when the 
particles in the pair are different and/or have different transverse 
momenta~\cite{Borghini:2004ra}, and is mostly due to anisotropic flow. 
On the other hand, momentum conservation affects the cosine 
term~(\ref{v2c(2p)}), which is non-zero even for particles with vanishing 
elliptic flow. 
Since $\bar v_2>0$ at relativistic energies, $v_{2,c}^{\rm pair}$ is positive 
for small-angle ($|\Delta\varphi_{12}|\leq\frac{\pi}{4}$) or large-angle 
($|\Delta\varphi_{12}|\geq\frac{3\pi}{4}$) pairs. 
This means that the yield of such pairs is larger in the reaction plane 
($\varphi^{\rm pair}\!\approx\Phi_R$) than perpendicular to it. 
Since $v_2(p_T)$ grows with increasing transverse momentum, as does the second
term in the right-hand of equation~(\ref{v2c(2p)}), $v_{2,c}^{\rm pair}$ increases 
with both ${p_T}_1$ and ${p_T}_2$, i.e.\ the anisotropy in the pair yield 
increases with the particle transverse momenta. 
Stated differently, for a pair of particles close in azimuth 
($\Delta\varphi_{12}$ close to 0), the anticorrelation~(\ref{f_c(2)}) is smaller 
(resp.\ larger) when the pair azimuth is along (resp.\ perpendicular to) 
$\Phi_R$, so that the pair yield is less (resp.\ more) ``suppressed'' by 
momentum conservation: there are overall more particles to balance the pair 
momentum along $\Phi_R$ than out-of-plane. 
Conversely, for a pair of back-to-back particles ($\Delta\varphi_{12}\approx\pi$)
momentum correlation induces a positive correlation~(\ref{f_c(2)}), which is 
larger if both particles lie along the reaction plane (resulting in 
$\varphi^{\rm pair}\approx\Phi_R\pm\frac{\pi}{2}$) than perpendicular to it. 

\begin{figure}[tp]
  \centerline{\includegraphics*[width=0.6667\linewidth]{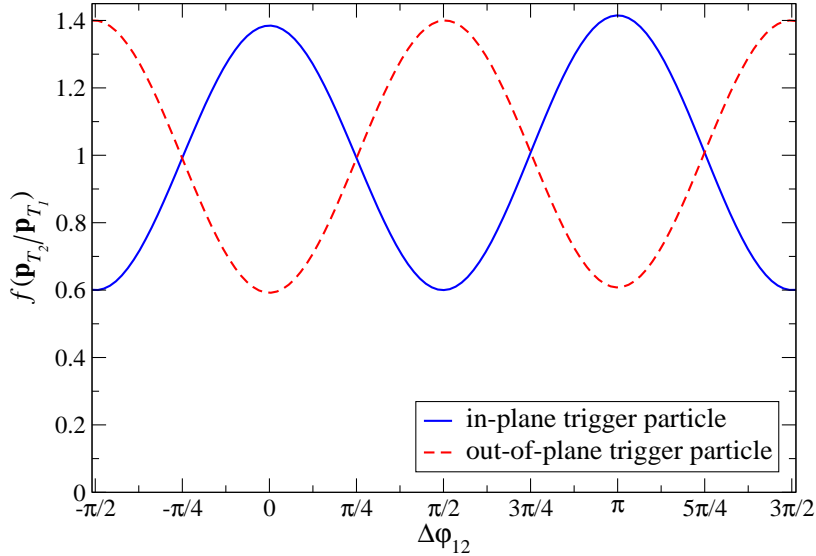}}
  \caption{\label{fig:f(2|1)}Conditional probability distribution 
  $f({{\bf p}_T}_2|{{\bf p}_T}_1)$ of the relative angle of an associated 
  particle for a trigger oriented along (full line) or perpendicular to (dashed 
  line) the reaction plane.}
\end{figure}

Rephrasing the above in yet another manner, one can investigate the 
{\em conditional\/} probability to find an ``associated'' particle (transverse  
momentum ${{\bf p}_T}_2$) given a ``trigger''particle (${{\bf p}_T}_1$), by 
dividing the pair distribution $f({{\bf p}_T}_1,{{\bf p}_T}_2)$ by the 
single-particle distribution $f({{\bf p}_T}_1)$. 
One then finds that for a trigger along the reaction plane 
($\varphi_1\approx\Phi_R$), there is a higher probability for associated 
particles close or away ($\Delta\varphi_{12}\approx 0$ or $\pi$) in azimuth than
around $\Delta\varphi_{12}\approx\pm\frac{\pi}{2}$. 
On the opposite, if the trigger escapes the system perpendicular to $\Phi_R$, 
the conditional probability to find an associated particle close or back-to-back
to the trigger ($\Delta\varphi_{12}\approx 0$ or $\pi$) is smallest. 
These rough features, entirely dictated by elliptic flow, are illustrated in 
figure~\ref{fig:f(2|1)}, where the values ${p_T}_1=6$~GeV/$c$,
${p_T}_2=4$~GeV/$c$, $v_2({p_T}_2)=0.2$, $\bar v_2=0.1$, 
$\mean{p_T^2}=(500$~MeV$/c)^2$ and $N=8000$ have been used. 
In further detail, for both in-plane or out-of-plane trigger particles, the 
probability for associated particles is higher away 
($\Delta\varphi_{12}\approx\pi$) than close ($\Delta\varphi_{12}\approx 0$) to 
the trigger: this is the trademark of the back-to-back correlation~(\ref{%
  f_c(2)}) induced by momentum conservation. 

\begin{figure}[tp]
  \centerline{\includegraphics*[width=0.6667\linewidth]{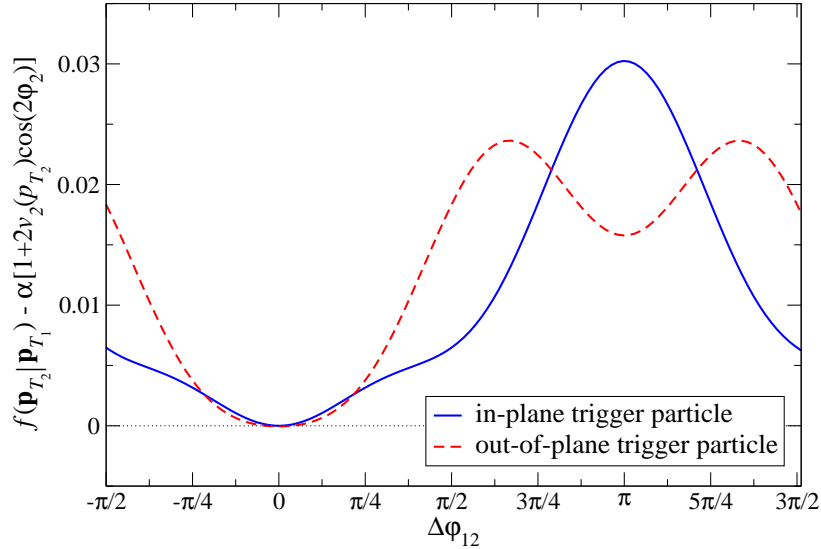}}
  \caption{\label{fig:f(2|1)-f(2)}``Flow-subtracted distribution of associated 
  particles'' $f({{\bf p}_T}_2|{{\bf p}_T}_1)-\alpha f({{\bf p}_T}_2)$ vs. the 
  relative angle, for a trigger particle along (full line) or perpendicular to 
  (dashed line) the reaction plane. 
  $\alpha$ is fixed by the ``ZYAM'' condition.}
\end{figure}

Now, the aim of a correlation study would be to investigate this finer 
structure. 
For that purpose, one would like to ``remove'' the anisotropic-flow-induced 
pattern. 
Given the general expression of the conditional probability distribution (resp.\
joint probability distribution) as a function of the two-particle 
cumulant~\cite{vanKampen}, the most natural --- although experimentally highly
challenging --- recipe would be to {\em divide\/} 
$f({{\bf p}_T}_2|{{\bf p}_T}_1)$ [resp.\ $f({{\bf p}_T}_1,{{\bf p}_T}_2)$] by the 
single-particle distribution $f({{\bf p}_T}_2)$ [resp.\ by the product 
$f({{\bf p}_T}_1)f({{\bf p}_T}_2)$], so as to isolate the correlation.\footnote{%
  This procedure does not entirely suppress the influence of the anisotropic 
  expansion, since the latter affects the strength of the correlation according 
  to the pair azimuth, see $\bar v_2$ in equation~(\ref{f_c(2)}).
  However, inasmuch as it is experimentally feasible, it would indeed cancel 
  the single-particle modulation induced by anisotropic flow, while leaving 
  intact the azimuthal dependence of the two-particle cumulant.} 
Yet the usual procedure is rather to {\em subtract\/} a flow-modulated 
background, whose normalization is for instance fixed by requiring that the 
yield vanish at its minimum (``ZYAM''~\cite{Ajitanand:2005jj}). 
Following this approach, the ``flow-subtracted distribution of associated 
particles'' is pictured in figure~\ref{fig:f(2|1)-f(2)} for both in- and 
out-of-plane trigger particles. 
As anticipated, this distribution is larger away from the trigger than close to 
it.
Furthermore, the away-side probability is smaller when the trigger points 
perpendicular to the reaction plane than when it points along $\Phi_R$: this 
was also not unexpected given the shape of the two-particle probability 
distribution~(\ref{f(2)}).
However, the detailed structures are quite non-trivial, in particular the dip 
at $\Delta\varphi_{12}=\pi$ in the case of an out-of-plane trigger, which both 
is a remnant of the ``incompletely'' subtracted anisotropic flow pattern.%
\footnote{If $f({{\bf p}_T}_2)$ had been divided from the conditional 
  probability, instead of being subtracted from it, the resulting quotient would
  have been a smooth first-harmonic sinusoid, rather than the curve in 
  figure~\ref{fig:f(2|1)-f(2)}.} 

The azimuthally-dependent behaviours described above follow solely from 
assuming 1.\ that the particle transverse momenta sum up to 0 and 2.\ that the 
particle emission is anisotropic: $\mean{p_x^2}>\mean{p_y^2}$. 
Those mere two ingredients are enough to give rise to non-trivial structures, 
which then have to be disentangled from those arising from additional physical 
effects. 
Two such patterns, whose analogues\footnote{The {\em qualitative\/} behaviours 
  discussed here are similar to those observed experimentally, but the 
  {\em quantitative\/} aspects differ.} have reportedly been observed at RHIC, 
were identified: 
(a) a lower away-side ($\Delta\varphi_{12}\approx\pi$) probability of associated 
particles in the case of an out-of-plane trigger compared to an in-plane 
trigger~\cite{Adams:2004wz}, and (b) a dip in the away-side yield of associated 
particles~\cite{Ulery:2005cc,Adare:2007vu}. 
These or similar structures have been predicted in models of parton energy loss 
that involve some path-length dependence~\cite{Wang:2000fq} [pattern (a)], or 
as signature of the interaction (``Mach cone'', ``gluon Bresstrahlung'', 
``Cerenkov ring'', ``jet deflection'', see e.g.\ reference~\cite{Adare:2007vu}) 
between the away-side parton and the medium through which it propagate 
[structure (b)].
If such models are to yield {\em quantitative\/} results, the possible 
contribution to the data of ``trivial'' correlations induced by 
momentum-conservation has to be investigated seriously.

\end{document}